\documentclass[twocolumn,prl,epsfig,showpacs,amsmath,amsfonts,amssymb,longbibliography]{revtex4-1}
\usepackage{graphicx}

\newcommand{\be}{\begin{eqnarray}}
\newcommand{\ee}{\end{eqnarray}}
\newcommand{\ra}{\rangle}
\newcommand{\la}{\langle}
\newcommand{\n}{\noindent}
\newcommand{\coll}{{\rm coll}}

\begin{document}

\title{Quantum Mechanics of Consecutive Measurements}
\author{Jennifer R. Glick}
\author{Christoph Adami}
\email[Corresponding author.\\]{adami@msu.edu}
\affiliation{Department of Physics and Astronomy, Michigan State University, East Lansing, Michigan 48824, USA}



\begin{abstract}Consecutive quantum measurements performed on the same system can reveal fundamental insights into quantum theory's causal structure, and probe different aspects of the quantum measurement problem. According to the Copenhagen interpretation, measurements affect the quantum system in such a way that the quantum superposition
collapses after the measurement, erasing any knowledge of the prior state. We show that a sequence of measurements in a collapse picture is equivalent to a quantum Markov chain, and that considering the unitary evolution of quantum wavefunctions interacting consecutively with more than two detectors reveals an experimentally measurable difference between a collapse and unitary picture. The non-Markovian nature of sequential measurements that we report is consistent with earlier discoveries in optimal quantum state discrimination.
\end{abstract}
\maketitle
{\em Introduction.}---The physics of consecutive (sequential) measurements on the same quantum system has enjoyed increased attention as of late, as it probes the causal structure of quantum mechanics~\cite{Brukner2014}. It is of interest to researchers concerned about the apparent lack of time-reversal invariance of Born's rule~\cite{Rovelli2015,OreshkovCerf2015}, as well as to those developing a consistent formulation of covariant quantum mechanics~\cite{ReisenbergerRovelli2002,OlsonDowling2007}, which does not allow for a time variable to define the order of (possibly non-commuting) projections~\cite{OreshkovCerf2014}.

Consecutive measurements can be seen to challenge our understanding of quantum theory in an altogether different manner, however. According to standard theory, a measurement causes the state of a quantum system to ``collapse'', re-preparing it as an eigenstate of the measured operator so that after multiple consecutive measurements on the quantum system any information about the initial preparation is erased. However, recent investigations of sequential measurements on a single quantum system with the purpose of optimal state discrimination have already hinted that quantum information survives the collapse~\cite{Nagalietal2012,Bergouetal2013}. In this Letter, we ask if it is possible to consistently assume that consecutive measurements form a Markov chain~\cite{HaydenMarkov2004,Datta2015}, that is, whether measurements ``wipe the slate clean".
While the suggestion that the relative state description of quantum measurement~\cite{Everett1957} (see also ~\cite{Zeh1973,Deutsch1984,CerfAdami1996,CerfAdami1998}) and the Copenhagen interpretation are at odds and may lead to measurable differences has been made before~\cite{Zeh1973,Deutsch1984}, here we frame the problem of consecutive measurements in the language of quantum information theory. The formulation we present suggests experimentally measurable differences between a collapse and unitary picture, in the density matrices and von Neumann entropies of the joint state of detectors. We will argue that a collapse picture of quantum measurement is therefore untenable for multiple consecutive measurements.

We start by carefully constructing the initial state as an arbitrary mixed quantum state $\rho_Q$. We can ``purify" $\rho_Q$ by defining a pure state where $Q$ is entangled with a reference system $R$~\cite{NielsenChuang_Book}
\be
|QR\ra=\sum_{n=1}^d\alpha_n|r_n\ra |n\ra_R\;. \label{state}
\ee
Here, $d$ is the dimension of the quantum state's Hilbert space, the $\alpha_n$ are arbitrary complex coefficients, and the states of $R$ are denoted by $|n\ra_R$ while those of $Q$ are labeled $|r_n\ra$. In what follows, we first assume that $Q$ is an ``unprepared'' or ``unknown'' state with maximum entropy so that in (\ref{state}) it is maximally entangled with $R$, i.e., $\alpha_n=1/\sqrt d$. With such an assumption, we do not bias any subsequent measurements~\cite{Wootters2006}.  We later discuss what happens in consecutive measurements on a known (that is, prepared) quantum state.

{\em Measurement of unprepared quantum states.}---To measure $Q$ with a detector (or ancilla) $A$~\footnote{We focus here on orthogonal measurements, a special case of the more general POVMs (positive operator-valued measures) that use non-orthogonal states. What follows can be extended to POVMs, while at the same time Neumark's theorem guarantees that any POVM can be realized by an orthogonal measurement in an extended Hilbert space.} 
we rewrite the quantum state in terms of the detector's eigenstates $|i\rangle_A$, which automatically serve as the ``interpretation basis"~\cite{Deutsch1984}, using the unitary matrix $V_{ni}=\langle a_i|r_n\rangle$ (letters  $a,b,\ldots$ with subscripts $i,j,\ldots$ indicate the basis $Q$ is written in, while $i,j,\ldots$ label the ancilla's basis states). We then entangle $Q$ with $A$, in initial state $|0\ra_A$, using a unitary entangling operation $\hat{U}_E$~\cite{CerfAdami1998} .
\be
|QRA \rangle = \hat{U}_E |QR\rangle |0\rangle_A = \frac{1}{\sqrt d} \sum_{ni} V_{ni} |a_i\rangle |n\rangle_R |i\rangle_A \;. ~~~
\ee
We can rewrite the reference's states
in terms of the $A$ basis by defining $|i\rangle_R = \sum_n V^\intercal_{in}|n\rangle_R$ with the transpose of $V$, so that the joint system $QRA$ appears as
(we omit from now on the labels from the ancilla states)
\be
|QRA\ra=\frac1{\sqrt d}\sum_i|a_i\ra |i\ra_R |i \ra\;. \label{schmidt}
\ee 
Tracing out the reference from the full density matrix $\rho_{QRA} = |QRA\ra\la QRA|$, we note that the detector is correlated with the quantum system~\cite{CerfAdami1998} and each has maximum entropy $S_Q \!=\! S_A \!=\! \log d$. The von Neumann entropy is defined as $S_X = S(\rho_X) = - \mathrm{Tr} (\rho_X \log \rho_X)$ for a density matrix $\rho_X$. We also note in passing that $R$ can be thought of as representing all previous measurements of the quantum system that have occurred before $A$.

We now measure $Q$ again, but in a rotated basis $|a_i\ra=\sum_jU_{ij}|b_j\ra$, by entangling it with an ancilla $B$. Unitarity implies that $\sum_j|U_{ij}|^2 \! = \! \sum_i|U_{ij}|^2 \! = \! 1$. Then, with $|j\ra$ the basis states of ancilla $B$ and $|ij\ra = |i\ra|j\ra$,
\be
|QRAB\ra=\frac1{\sqrt d}\sum_{ij}U_{ij}|b_j\ra |i\ra_R |ij\ra \;.
\ee  
It is easy to show that $\rho_A \!=\! \rho_B \!=\! 1/d\sum_i | i\rangle\langle  i |$, so that both detectors have maximum entropy $\log d$, while
\be
\rho_{AB}=\frac1d\sum_{i}|i\ra\la i| \otimes \sum_j |U_{ij}|^2 |j\rangle\langle j|  \;. \label{rhoab}
\ee
Equation~\eqref{rhoab} immediately implies that if the quantum system is measured repeatedly in the same basis ($U_{ij} = \delta_{ij}$) by independent detectors, all of those detectors will be perfectly correlated (they reflect the same outcome), creating the illusion of a wavefunction collapse~\cite{CerfAdami1996,CerfAdami1998}. 

The preceding results are entirely consistent with the standard formalism for orthogonal measurements~\cite{Peres1995,Holevo2011}, where the conditional probability $p_{j|i}$ to observe outcome $j$, given that the previous measurement yielded outcome $i$, is
\be
p_{j|i}=|U_{ij}|^2\;. \label{cond}
\ee
Indeed, our findings thus far are fully consistent with a picture in which a measurement collapses the quantum state (or alternatively, where a measurement recalibrates an observer's ``catalogue of expectations"~\cite{Schroedinger1935,Englert2013,Fuchsetal2014}). To see this, 
we write the joint density matrix of detector $A$ that records outcome $i$ with probability $1/d$ and detector $B$ that measures the same quantum state at an angle determined by the rotation $U$
\be
\rho^\coll_{AB}=\frac1d\sum_i |i\rangle \langle i| \otimes \rho^i_B ~ ,
\ee
with $\rho^i_B$ defined using projection operators $P_i = |i\ra\la i|$
\be
\rho^i_B=\frac{{\rm Tr}_A\left(P_i\rho_{AB}P_i^\dagger\right)}{ {\rm Tr}_{AB}\left(P_i\rho_{AB}P_i^\dagger\right) }=\sum_j |U_{ij}|^2|j\rangle\langle j|\;. \label{proj}
\ee

Let us perform another measurement of the quantum system using an ancilla $C$ such that $|b_j\ra=\sum_kU'_{jk}|c_k\ra$.
We then find (here and before, indices $i$ refer to $A$, $j$ to $B$, and now $k$ to $C$)
\be
|QRABC\ra=\frac1{\sqrt d}\sum_{ijk}U_{ij}U'_{jk}|c_k\rangle |i\rangle_R |ijk\ra\;. \label{qabc}
\ee
Tracing out $Q$ and $R$ from the density matrix $\rho_{QRABC}$, as we do not observe either the quantum system nor reference, leads to the joint state of three detectors
\begin{equation}
\rho_{ABC}= \! \frac1d \! \sum_{i}|i\ra\la i| \otimes \sum_{jj'} U_{ij}U^*_{ij'}|j\ra\la j'|\otimes\sum_kU'_{jk}U^{\prime *}_{j'k}|k\ra\la k| \: .\label{full}
\end{equation}
Tracing this expression over $C$ recovers $\rho_{AB}$ in Eq.~(\ref{rhoab}) as it should because the measurement $C$ does not affect the joint state of the past measurements $A$ and $B$. Tracing over $B$ gives
\be
\rho_{AC}=\frac1d \sum_{i} |i\ra\la i|\otimes\sum_{jk}|U_{ij}|^2 \, |U'_{jk}|^2 \, |k\ra\la k|\;, \label{ac}
\ee
while tracing over $A$ yields
\be
\rho_{BC}=\frac1d \sum_{j} |j\ra\la j|\otimes\sum_k|U'_{jk}|^2 \, |k\ra\la k|\;. \label{bc}
\ee

All three pairwise density matrices are diagonal in the detector product basis (see Theorem 1 in the Supplementary Material~\cite{supp}). We can take ``diagonal in the detector product basis" to be synonymous with ``classical". At the same time, each detector has classical information about the quantum system
(Theorem 2 in Supplementary Material~\cite{supp}).
Furthermore, all pairwise density matrices discussed here, and their corresponding entropies, are identical to those in a collapse picture. 

However, the joint state of all three detectors when assuming a collapse picture is incoherent
\begin{equation}\label{coll}
\rho_{ABC}^\coll \!=\! \frac1d \! \sum_{i} \! |i\ra\la i|  \otimes \! \sum_j \! |U_{ij}|^2|j\ra\la j|  \otimes \! \sum_k \! |U'_{jk}|^2|k\ra\la k|, 
\end{equation}
unlike expression~(\ref{full}) obtained in the unitary picture, which is coherent due to the non-diagonality of the $B$ subsystem. The presence of these additional terms in~(\ref{full}) has fundamental consequences for our understanding of the measurement process. After all, the three measurements were implemented as projective measurements, which according to the traditional view
``reduce" the wavefunction of the system. Indeed, such an apparent collapse has taken place after the second consecutive measurement (\ref{rhoab}) as the corresponding density matrix has no off-diagonal terms. However, the third measurement seemingly {\em undoes} this projection, as can be seen from the appearance of off-diagonal terms in~\eqref{full}. This ``reversal" is different from protocols that can ``un-collapse" weak measurements~\cite{korotkov2006,Jordan2010}, because it is clear that the wavefunction~\eqref{qabc} underlying the density matrix is and remains unprojected. 

\setlength\belowcaptionskip{-1ex}
\begin{figure}[t]
 \includegraphics[width=1.0\linewidth]{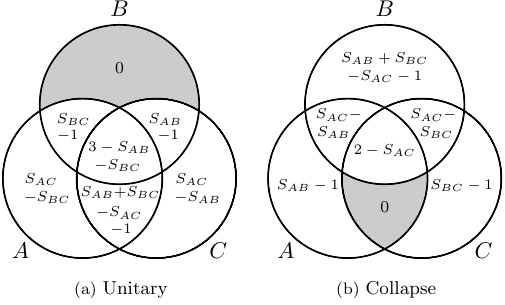}
   \caption{Entropy Venn diagram for the joint state of three qudit detectors $A$, $B$, $C$ that consecutively measure an unprepared quantum system $Q$ (entropies with logarithm to base $d$). (a) A unitary description of quantum measurement where the middle detector $B$ is fully known given the past $A$ and future $C$ (grey shaded area). (b) A collapse description, where $A$ and $C$ are independent when given $B$ (grey shaded area).}
\label{fig:venn1}
\end{figure}
To determine if the reported survival
of the quantum superposition has measurable consequences, we calculate the entropy of each pair of detectors and of the joint state of all three detectors. Taking logarithms to base $d$, the pairwise entropies follow directly from~\eqref{rhoab},~\eqref{bc}, and~\eqref{ac}:
\be\label{pair_entropies}
S_{AB} \! & = & \! 1 \! - \frac{1}{d} \sum_{ij} |U_{ij}|^2 \log |U_{ij}|^2,  \label{SAB}\\
S_{BC} \! & = & \! 1 \! - \frac{1}{d} \sum_{jk} |U'_{jk}|^2 \log |U'_{jk}|^2, \label{SBC}\\
S_{AC} \! & = & \! 1 \! - \frac{1}{d} \sum_{ik} \! \Big( \! \sum_j |U_{ij}|^2  |U'_{jk}|^2 \Big) \! \log \! \Big( \! \sum_{j'} |U_{ij'}|^2  |U'_{j'k}|^2 \Big). \nonumber \\ \label{SAC}
\ee
Furthermore, $S_{AC}$ is equal to $S_{ABC}$, the entropy of $\rho_{ABC}$ (this holds for any three consecutive detectors, see Corollary 1.1 in Supplementary Material~\cite{supp}). From the definition of conditional entropy~\cite{CerfAdami_PRL1997} it follows that, given the measurement $A$ in the past and the measurement $C$ in the future, detector $B$'s state is fully determined
(see grey area in Fig.~\ref{fig:venn1}(a))
\be
S(B|AC) = S_{ABC} - S_{AC} = 0 \: .
\ee
It can be shown quite generally that this quantity does not vanish in a collapse picture (Corollary 1.1 in Supplementary Material~\cite{supp}). 

We now briefly show that the measurement chain in a collapse picture is Markovian, as defined in~\cite{HaydenMarkov2004} (see also~\cite{Datta2015} and references therein). From~\eqref{coll}, the joint entropy of all three detectors (using $H$ to distinguish collapse entropies from $S$, the entropies in the unitary picture) is
\begin{equation}\label{entropy_ABC_collapse}
H_{ABC} = 1 \! -  \frac{1}{d} \! \sum_{ij} \! |U_{ij}|^2 \! \log |U_{ij}|^2 - \frac{1}{d} \! \sum_{jk} \! |U'_{jk}|^2 \! \log |U'_{jk}|^2 ,
\end{equation}
or, $H_{ABC} = S_{A} + H(B|A) + H(C|B)$. Using the chain rule for entropies~\cite{CerfAdami1998}, $H_{ABC} = S_A + H(B|A) + H(C|BA)$, we see immediately that $H(C|BA) = H(C|B)$, the Markov property for entropies~\cite{HaydenMarkov2004,Datta2015}. 
This further implies that subsystems $A$ and $C$ 
are independent from the perspective of $B$, since the conditional mutual entropy~\cite{CerfAdami1998} vanishes (see grey area in Fig.~\ref{fig:venn1}(b))
\be 
 H(A:C|B) & = & H(C|B) - H(C|BA) = 0 ~ .
\ee 
\n The equivalent quantity does not vanish in the unitary formalism, reflecting the fundamentally non-Markovian nature of the quantum chain of measurements (see Theorem 3 in Supplementary Material~\cite{supp} for a derivation in an arbitrarily long chain). Figure~\ref{fig:venn1} uses quantum entropy Venn diagrams (see, e.g.,~\cite{CerfAdami1998}) to highlight these key differences between the two pictures.

\begin{figure}[t]
 \includegraphics[width=0.9\linewidth]{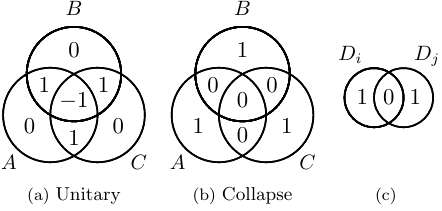}
   \caption{Entropy Venn diagram for the joint state of three qubit detectors $A$, $B$, $C$. Detector $B$ measures $Q$ at an angle $\theta=\pi/4$ relative to the basis of $A$, and $C$ measures at $\theta'=\pi/4$ relative to the basis of $B$. Venn diagram based on (a) unitary evolution of the wavefunction, and (b)  according to the collapse picture. (c) In both cases any two detectors $D_i$ and $D_j$ are uncorrelated when the third is traced out.}
\label{fig:venn2}
\end{figure}
We can readily apply this formalism to the specific case of qubits ($d \!=\! 2$). Measurements with detector $B$ at an angle $\theta$ relative to the previous measurement $A$, and $C$ at an angle $\theta'$ to $B$, can each, without loss of generality, be implemented with a rotation matrix of the form  
\be
U&=&\left( \! \begin{array}{cc}\cos(\theta)& -\sin(\theta)\nonumber\\
                           \sin(\theta)&  \;\;\; \cos(\theta)\nonumber
              \end{array} \! \right) \: .
\ee
For measurements at $\theta = \theta' =\pi/4$ for example, we have $|U_{ij}|^2=|U^\prime_{ij}|^2=1/2$, and we expect the outcome of each measurement to be random, that is, $S_A \! = \! S_B \! = \! S_C \! = \! 1$ bit. The joint entropy of each pair of detectors is two bits, as can be read off of Eqs.~(\ref{SAB}-\ref{SAC}). Because of the non-diagonal nature of (\ref{full}), the joint density matrix of the three detectors (using $\sigma_z$, the third Pauli matrix, and $\mathbb{I}$, the identity of dimension 2)
\be  
\rho_{ABC} = \frac{1}{8} \label{three}
\begin{pmatrix}
\mathbb{I}  & -\sigma_z  & 0           & 0          \\
-\sigma_z   & \mathbb{I} & 0           & 0          \\ 
0           & 0          & \mathbb{I}  & \sigma_z   \\
0           & 0          & \sigma_z    & \mathbb{I} 
\end{pmatrix} , 
\ee
has entropy $S_{ABC}=2$ bits, as can be checked by finding the eigenvalues of (\ref{three}). The collapse density matrix (\ref{coll}) on the other hand gives $H_{ABC}=3$ bits, as can be verified from~\eqref{entropy_ABC_collapse}. Figure~\ref{fig:venn2} summarizes the entropic relationships for qubits in the two pictures. 

It is instructive to note that the Venn diagram in Fig.~\ref{fig:venn2}(a) is the same as the one obtained for a one-time binary cryptographic pad where two classical binary variables (the source and the key) are combined to a third (the message) via a controlled-NOT operation~\cite{Schneidman2003} (the density matrices underlying the Venn diagrams are very different, however). Still, it implies that the state of any one of the three detectors can be predicted from knowing the joint state of the two others, in clear violation of the collapse postulate that a measurement wipes clean the history of the quantum state. 
However, the prediction of $C$ cannot be achieved using expectation values from $B$'s and $A$'s states separately, as the diagonal of (\ref{full}) corresponds to a uniform probability distribution. For the qubit case, the difference between the density matrix $\rho_{ABC}$ and the collapse version can be ascertained by revealing the off-diagonal terms via quantum state tomography (see, e.g.,~\cite{Whiteetal1999}), or by measuring just a single moment~\cite{Tanaka2014} of the density matrix, such as ${\rm Tr} (\rho_{ABC}^2)$. 

{\em Measurement of prepared quantum states.}---Suppose a quantum system is prepared in the known state
\be
\rho_Q=\sum_{j=1}^d p_j |a_j\ra\la a_j|\;, \label{prep}
\ee
which we already wrote in the basis of ancilla $A$, as this will be the first measurement. We can always prepare a state like~\eqref{prep} by measuring an unknown quantum state in a given, but arbitrary, basis. Then, a second measurement at a relative angle $\theta$ gives rise to a projected state for $Q$ that is equivalent to the density matrix (\ref{proj}). If we choose for the state preparation only the outcome $i=0$, for example, then $p_j=|U_{0j}|^2$ provides the probability distribution.

The purification of (\ref{prep}) in terms of $A$'s basis is 
\be\label{rhoa}
|QA\ra=\sum_{i}\sqrt p_i|a_i\ra |i\ra\;, 
\ee 
creating a correlated state with $S_Q \!=\! S_A \!=\! -\sum_i p_i\log p_i$. We now proceed as before. Introduce detector $B$ with its eigenbasis $\la b_j|a_i\ra=U_{ij}$ (this $U$ is not to be confused with $U$ in the definition of $p_j$ above). After entanglement,
\be
|QAB\ra=\sum_{ij}\sqrt{p_i} ~ U_{ij}|b_j\ra |ij\ra \: , 
\ee 
giving rise to 
\be
\rho_{AB}&=&\sum_{ii'}\sqrt{ p_i p_{i'}} ~ |i\ra\la i'|\otimes \sum_j U_{ij} U^*_{i'j} |j\ra\la j|\;, \\
\rho_{B}&=&\sum_{ij}p_i|U_{ij}|^2 |j\ra\la j|\;. \label{rhob}
\ee
The entropy of $B$ is naturally $S_B=-\sum_j q_j\log q_j$, where $q_j=\sum_i p_i |U_{ij}|^2$ is the marginal probability obtained from the joint probability $p_{ij}=p_i \: p_{j|i}$. The conditional probability $p_{j|i}$ of obtaining outcome $j$ with $B$, given that we had obtained outcome $i$ with $A$, was defined in (\ref{cond}). Introducing detector $C$, the joint density matrix for all three detectors is ($\rho^\coll_{ABC}$ has no such off-diagonal terms)
\begin{equation}\begin{split}
\rho_{ABC}=\sum_{ii'}\sqrt{p_i p_{i'}} |i\ra\la i'| & \otimes \sum_{jj'} U_{ij}U^*_{i'j'}|j\ra\la j'|  \\
                                                    & \otimes\sum_k U'_{jk} U^{\prime *}_{j'k}|k\ra\la k| \:. \label{full=k}
\end{split}\end{equation}
From~\eqref{full=k}, detector $C$'s entropy is $S_C = - \sum_k q'_k \log q'_k$, where $q'_k = \sum_{ij} p_i \: |U_{ij}|^2 \: |U'_{jk}|^2$ is obtained from the joint probability $p_{ijk}=p_i \: p_{j|i} \: p_{k|j}$, and $p_{k|j}=|U'_{jk}|^2$.  

\begin{figure}[t]
\begin{center}
 \includegraphics[width=0.7\linewidth]{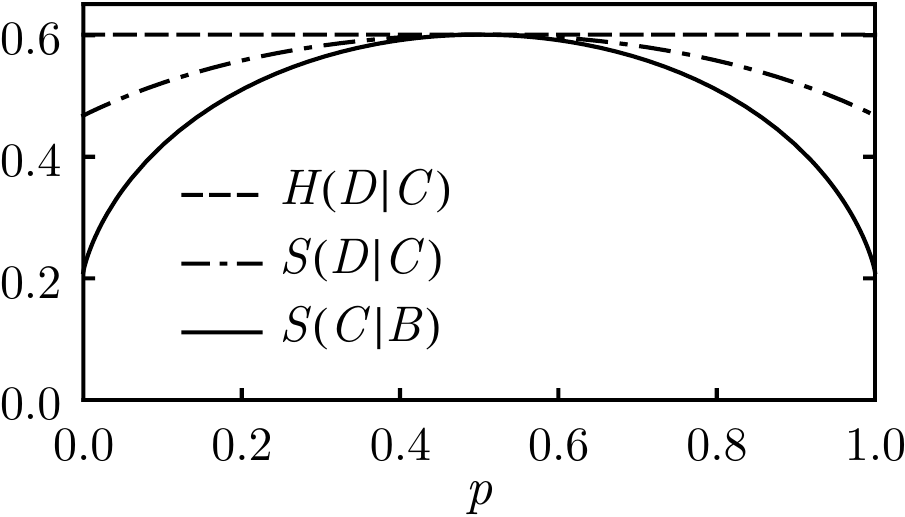} ~~~
\end{center}
   \vspace*{-5mm}
   \caption{Conditional entropies $S(D|C)$ and $S(C|B)$ in the unitary picture and $H(D|C)$ in the collapse picture, for three consecutive measurements on the prepared state~\eqref{prep} as a function of the state preparation $p$. Each detector is at an angle $\pi/8$ relative to the previous detector. For these angles, $H(D|C)=H(C|B)$.
   }
\label{fig:qubits}
\end{figure}
We apply these results once more to qubits,
with three measurements $B$, $C$, and $D$ of the quantum system after the preparation with $A$, each at an angle $\pi/8$. In Fig.~\ref{fig:qubits}, we show that
in a unitary description all measurements prior to $C$ leave a trace: the conditional entropies $S(C|B)$ and $S(D|C)$ retain a dependence on the preparation $p$. The collapse entropy $H(D|C)$, on the contrary, is independent of $p$. This formalism can also be used to succinctly describe the quantum Zeno~\cite{HomeWhitaker1997,Peres1995} and anti-Zeno~\cite{KaulakysGontis1997,LewensteinRzazewski2000,Luis2003} effects (see Supplementary Material~\cite{supp}).

{\em Conclusions.}---Conventional wisdom in quantum mechanics dictates that the measurement process ``collapses'' the state of a quantum system so that the probability a particular detector fires depends only on the state preparation and the measurement chosen. 
Using a quantum-information-theoretic approach, 
we have argued that a collapse picture makes predictions that differ from those of the unitary (relative state) approach if multiple consecutive measurements are considered. Should future experiments corroborate the manifestly unitary formulation we have outlined, such results would further support the notion of the reality of the quantum state~\cite{Pusey2012} and that the wavefunction is not merely a bookkeeping device that summarizes an observer's knowledge about the system~\cite{Englert2013,Fuchsetal2014}. We hope that moving discussions about the nature of quantum reality from philosophy into the empirical realm will ultimately lead to a more complete (and satisfying) understanding of quantum physics.

\begin{acknowledgments}
CA would like to thank N. J. Cerf and S. J. Olson for discussions, and acknowledges support by the Army Research Offices grant \# DAAD19-03-1-0207. Financial support by a Michigan State University fellowship to JRG is gratefully acknowledged.
\end{acknowledgments}

%

\end{document}